\documentclass[conference, 12pt]{IEEEtran}
\usepackage{makeidx}  
\usepackage{cite}
\usepackage{graphicx}
\usepackage{mdwtab}
\usepackage{algorithm}
\usepackage{algorithmic}
\usepackage{amsmath}
\usepackage{tabularx}
\usepackage{dirtree}
\usepackage{tikz}

\usepackage[a4paper,left=15mm,right=15mm, top=1cm, bottom=2cm]{geometry}

\begin{document}
	 
\title{Intelligent Reflecting Surface Aided Wireless Energy Transfer and Mobile Edge Computing for Public Transport Vehicles}

\title{%
  Intelligent Reflecting Surface Aided \\ Wireless Energy Transfer and Mobile Edge Computing for Public Transport Vehicles \\ --
  \Large \\ A Communication Eco-System For Transport Vehicles of 6G-era.}

\author{\IEEEauthorblockN{Shan Jaffry}
	\IEEEauthorblockA{
		School of Internet of Things, \\ Xi'an Jiaotong-Liverpool University.\\
		shan.jaffry@xjtlu.edu.cn
		}
}
	
	\maketitle
	\begin{abstract}
    In the forthcoming 6G era, Future Public transport vehicles (F-PTV), such as buses, trains etc. will be designed to cater the communication needs of the commuters that will carry numerous smart devices (smartphones, e-bands etc.). These battery-powered devices need frequent recharging. Since recharging facilities are not readily available while commuting, we envision F-PTVs that will provide in-vehicle recharging via Wireless Energy Transfer (WET) through in-vehicle Access Points. F-PTV will also be internally coated with Intelligent Reflecting Surface (IRS) that reflect incident radio waves towards the intended device to improve signal strength at the receiver, for both information and energy transmissions. F-PTVs will also be equipped with Mobile Edge Computing (MEC) servers to also serve multiple purposes, including reduction in devices’ task completion latency and offering in-vehicle Cloud services. MEC-server offloading will also relieve smart devices' processors off intensive tasks to preserve battery power. The challenges associated with the IRS-aided integrated MEC-WET model within F-PTV are also discussed. We present simulation to show the effectiveness of such a model for F-PTV.
	\end{abstract}
	
	
	
	\section{Introduction}
  
    Modern applications running on smart devices and gadgets are increasingly becoming sophisticated and consume excessive computation resources and battery power. On the other hand, the available computational and battery resources are limited for such devices, mainly due to their small form factor. Replenishing devices' battery levels on-the-go is becoming a serious challenge. In particular, mobile users that are on the move, for example, commuters inside public transport vehicles (PTVs), such as buses, trains, subways etc., are more vulnerable to battery drainage as they have limited recharging facilities available. These users collectively carry a large number of  devices and engage in online or offline activities during the commute. Such activities (e.g. audio/video calling, gaming, web-surfing, etc.)  drain extensive amount of devices' battery.
    
    On the other hand, it has been proposed that future PTVs (F-PTVs) will have a dedicated in-vehicle small cell to provide services to commuting users \cite{jaffry2016making}. Primarily, Access Points within F-PTV (PTV-AP) are aimed to enhance in-vehicle cellular user's quality of service (QoS). The commuters within F-PTV will be effectively decoupled from the out-of-vehicle macro-cell base station which significantly reduce cellular network load. Furthermore, F-PTV will also reduce the group handover into a collective hand-off by the vehicle \cite{jaffry2020comprehensive}. Integrating in-vehicle wireless energy transfer (WET) methods with PTV-APs can enable wireless charging of commuters’ energy-constrained devices and gadgets while on the move.

    However, due to hostility of air interface towards electromagnetic radio signals, the magnitude of power delivered though WET remains effectively very low. To make matter worse, signal quality degrades drastically in a crowded public transport as signal passes through a human bodies. This body-shadowing effect further diminishes the prospect of any energy gains with WET-enabled PTV-AP. To bolster wireless channels within F-PTV, we further envision installing the Intelligent Reflecting Surface (IRS) within the vehicle. The IRS are passive reflecting elements that change the phase and amplitude of wireless signals and can effectively undo the signal attenuation by reflecting incident waves towards the receiver device. 
    	
    \begin{figure*}[!t]\centering 
	\includegraphics[width= 5.5 in, trim={0cm 0cm 0cm 0.0cm},clip = true]{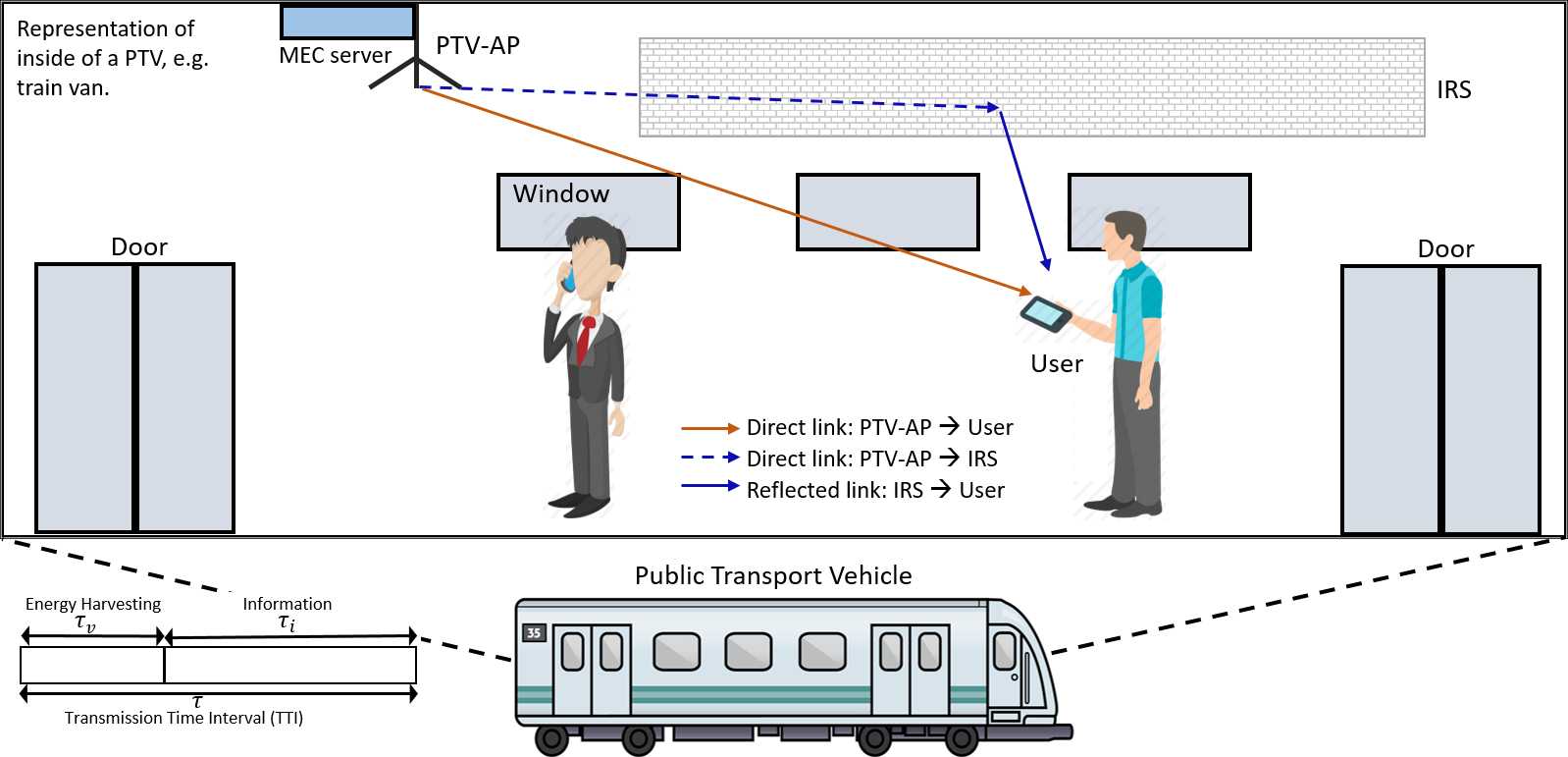}
	\caption{A depiction of a communication-friendly futuristic Public Transport Vehicle for 6G era(An idea).}
	\label{fig_1}
    \end{figure*} 
 
    On the other hand, to reduce increasing computing requirements from the smart devices, researchers have focused on offloading services and tasks to the near-by external servers which are relatively more powerful than smart devices and gadgets. These powerful computing nodes, also known as MEC-servers, are located at the network edge, for example with the base-stations and APs. These MEC-servers can be collocated with PTV-AP to provide in-vehicle Cloud services and augment computing requirements of commuting users inside F-PTV via task offloading. The task offloading can be exploited for dual, albeit closely linked, purposes. First, offloading reduce tasks' processing time as MEC-server’s Central Processing Units (CPUs) are more powerful than smart devices. Second, the task offloading also preserves extensive amount of device's power that is otherwise consumed during the local computations. 

    This envisioned IRS-aided MEC-WET system creates a cellular-friendly eco-system within F-PTV in the 6G era. In the rest of this article, we first provide a brief overview of the envisioned combined IRS-aided integrated MEC-WET system for F-PTVs. We then discuss potential challenges and possible solutions associated with the envisioned F-PTV. Later, Simulation and numerical results are provided to validate the effectiveness of the design.

    \section{IRS-Aided MEC-WET within Public Transport Vehicle}
    In this section we will first discuss about the WET technology followed by IRS and MEC system within F-PTV.
   
    \subsection{Suitable WET Technology for PTV}
    
    WET is not an entirely new idea. In fact, it is more than century old concept \cite{shinohara2011power}. Famous inventor Nikola Tesla conducted wireless power transfer experiments in the late 19th century \cite{tesla1904transmission}. However, modern improvements in microelectronic circuit design have sparked a new interest in WET for smart devices, gadgets, and sensor networks.   
    
   We broadly categorized WET technologies into two types. The first type, called Inductive Coupling WET (IC-WET), exploits near-field electromagnetic energy transmission phenomena. IC-WET is popular for charging smart gadgets, electric vehicles, implanted medical devices etc. More recently, key smart device manufacturers have built practical wireless charging accessories based on IC-WET. However, IC-WET works only for very short distances and in most cases the charging adapter and the rechargeable device needs to be within near-physical contact. Due to this limitation, IC-WET is not a suitable fit for F-PTV use case.
   
    The second type of WET uses radio frequency (RF) signals to transmit energy and offers flexibility of charging devices at a relatively larger distances (few meters). This makes RF-WET a more suitable option for the use case within F-PTV. However, electromagnetically radiated RF signals withers significantly in-air before reaching the receiver. The highly attenuated signal received at the harvesting device is significantly low, and even ineffective at times, to provide any useful energy gains. The signal attenuation exacerbates within F-PTV because of human-body shadowing due to presence of passengers.

    \subsection{IRS Coating within Vehicle} 
    
        The poor wireless propagation environment within the F-PTV can be alleviated by internally coating passive IRS elements within the vehicle. An IRS can virtually reconfigure RF propagation environment by controlling the phase and amplitude of incident waves \cite{basar2019wireless}. Interestingly, unlike legacy relaying systems, IRS do not require external power source to boost signal strength. Instead, IRS comprise of array of low-cost passive reflecting elements that can be controlled in real-time to focus the direction of transmitted wave towards the receiving device, much like a mirror. An in-vehicle IRS controller is responsible for adjusting amplitude and phase of incident waves. The waves can be added constructively to significantly improve the signal strength at the desired receiver. The IRS-aided constructive interference can be achieved both for information and energy harvesting signals. The same mechanism can be used to induce destructive interference at unwanted or eavesdropping receivers to nullify the undesirable interference effect. 
 
        Due to passive nature of IRS elements, additional power for signal decoding or amplification is not required. Furthermore, unlike active relays IRS also do not impose additional white noise as a consequence of decoding or amplification \cite{pan2020intelligent}. The IRS's RF reflecting elements can be conveniently coated over the existing structures such as on the walls or windows of the vehicle which significantly reduces the  deployment complexity \cite{basar2019wireless}.

        To fully exploit IRS benefits within F-PTV, controlling reflective elements is a challenging task. The IRS controller must have real-time knowledge of factors such as channel state information (CSI), number of wireless links, optimal phase and amplitude etc \cite{wu2019towards}. However, even a fixed or sub-optimal tuning of reflectors have shown to significantly improve the quality of signal-to-noise-ratio (SNR) or harvested energy at the receiver \cite{bai2020latency, pan2020intelligent}.

        Despite IRS-aided improved channel conditions, effective energy harvesting gain for the user devices remain limited due to low RF-to-DC power conversion efficiency and other hardware constraints. Hence, next we discuss MEC offloading to further improve energy preservation within F-PTV.

    \subsection{Edge Computing for Energy Preservation}
    
    When it comes to WET, RF-to-DC power conversion efficiency is practically much lower than theoretically anticipated  \cite{almansouri2018cmos}. This is in part true due to device's size limitation and hardware energy conversion constraints. Hence only limited energy gain could be achieved by relying solely on WET, even after IRS installation. On the other hand, a significant amount of devices' energy consumption is due to processing requirements of applications' tasks. This is true because a device's CPU’s power consumption is proportional to the operating frequency and the operational voltage of the CPU. By frequently forcing device's CPU into idle mode, significant energy savings can be achieved. One way of relieving device's CPU off continuous operations is by offloading tasks to the MEC-servers collocated with PTV-AP. This MEC offloading serves two purposes. Firstly, it allows faster completion of tasks as CPU(s) at MEC-servers are more powerful than user devices. Secondly, it preserves a significant amount of the device's battery power, courtesy to task offloading.  Furthermore in \cite{bai2020latency} researchers have shown that IRS-aided wireless environment also reduce system-level latency for MEC systems by improving SNR at the end receivers and also overall improves energy efficiency.  
    
    The IRS-aided MEC-WET system within F-PTV does not come without challenges which are discussed next.
 
    \section{Design Challenges and Future Research Directions}
    
    In this section we discuss some of the challenges associated with envisioned in-vehicle IRS-aided MEC-WET system.
    
    \subsection{Crowded Channels within PTV}
    
    Due to relatively higher device-density (devices per square meter) within public transport and installation of reflectors within F-PTV, in-vehicle radio channels may get over-crowded in case excessive information and energy signals are exchanged between PTV-AP and commuters. Even in the absence of online activity, users may compete for channels to harvest RF energy or to upload and download tasks to and from the MEC-servers, respectively. Given the insufficient number of available channels for both information and energy transmission, fair channel scheduling will be challenging for IRS aided environment within PTV. 
        
    Alternatively, the mmWave band, which is suitable for indoor communication, offers wider bandwidth and more radio channels to alleviate congestion problems. However, due to large propagation losses, mmWave band may not be suitable for energy signal transmission. One possible solution to resolve above stated issues is to use a combination of mmWave and sub-6 GHz (e.g. 2.4 GHz and 3.5 GHz) bands within F-PTV such that information signals are carried over mmWave frequencies and energy harvesting is carried over sub-6 GHz band.

    \subsection{Limited MEC-WET Resources within PTV}
    
    As compared to Cloud storage, MEC servers have relatively limited storage and computing resources. In particular, MEC servers within PTVs will have further reduced capabilities due to physical size limitations. Hence, optimally distributing the limited computing resources among several in-vehicle devices would be another challenge.  
 
    Furthermore, energy signal transmission will base on time-sharing scheme. For a given Transmission Time Internal (TTI) $\tau$, energy signal is transmitted for the period $\tau_v = \gamma \tau$, where $\gamma \in [0,1]$. The remaining time $\tau_i = (1-\gamma)\tau$, is dedicated for information transmission ($\tau_i > \tau_v$). Given that $\tau_v$ is the small proportion of $\tau$, allocating slots for numerous devices need efficient time scheduling. PTV-APs with Multiple-Input-Multiple-Output (MIMO) antenna may allow simultaneous transmission of multiple energy signals via spatial diversity. However,  scheduling energy resources for large number of users is still a challenge.  
    
\begin{figure}[!t]\centering
	\includegraphics[width=\linewidth, trim={0cm 0.1cm 0cm 0.0cm},clip = true]{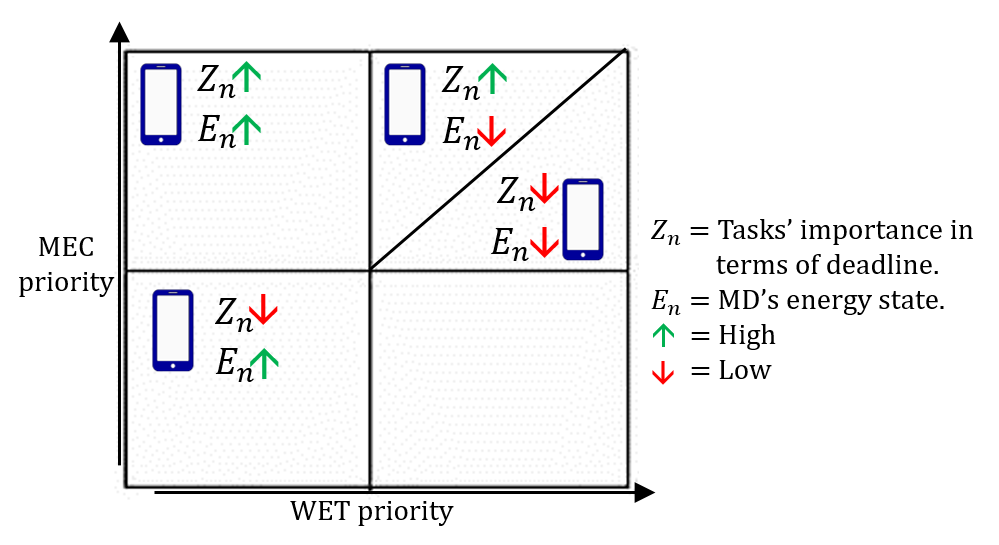}
	\caption{A simplistic decision grid for scheduling computing and energy resources scheduling among devices.}
	\label{fig_2b}
    \end{figure}

    To address computing and energy resource management and scheduling, smart optimization techniques are needed. One simplistic scheme could be to rely on the tasks' completion deadline ($Z_n$) and the battery level ($E_n$) for each device. Based on this information, PTV-AP can allocate MEC and WET resources for individual user. For example, as shown in Figure \ref{fig_2b} if a device has high battery level and lower deadline constraint, then that the device task can be performed locally. On the contrary, if a device has low battery levels then it is given higher priority for both MEC and WET resource allocation.  

    \subsection{Active and Passive Beamforming} 

    The internal structure of most public transport is generally suitable for coating IRS elements. For example, the roof or the upper sidewalls could be covered with the reflecting elements. Given the passive nature of IRS elements additional power is not required to improve signal quality as the IRS controller adjust the phase and amplitude of incident waves striking the meta-material surface. The installation of PTV-AP and IRS would follow Line-of-Sight (LoS) link. Active beamforming at the PTV-AP and passive beamforming at the IRS elements would guide the direction of the traveling waves. As mentioned earlier, the presence of large number of passengers would increase body-shadowing effect within PTV. In the downlink direction, the combined active and passive beamforming can alleviate this effect. However, active beamforming may not always be possible for the uplink transmission due to device (such as smartphones, watches, e-bands etc.) limitations. Still, researchers in \cite{pan2020intelligent,zhou2020computation} have shown that in an IRS aided environment, even the signals transmitted in random direction provide better quality than non-IRS environment.

    \subsection{Electromagnetic Radiation Concerns}
    Regular in-vehicle transmissions and additional IRS reflections inside public transport increase electromagnetic radiation inside the vehicle which raises health concerns for passengers. Hence, these concerns must be addressed by suitable researchers in line of standardized guidelines outlined by the governing bodies. 
    
    To demonstrate the feasibility of IRS-aided MEC-WET installation within PTV, numerical results are presented in the next Section. 
    
    \section{Simulation Results and Empirical Observation}
    
\begin{figure}[!t]\centering
	\includegraphics[width=\linewidth, trim={0cm 0cm 0cm 0.0cm},clip = true]{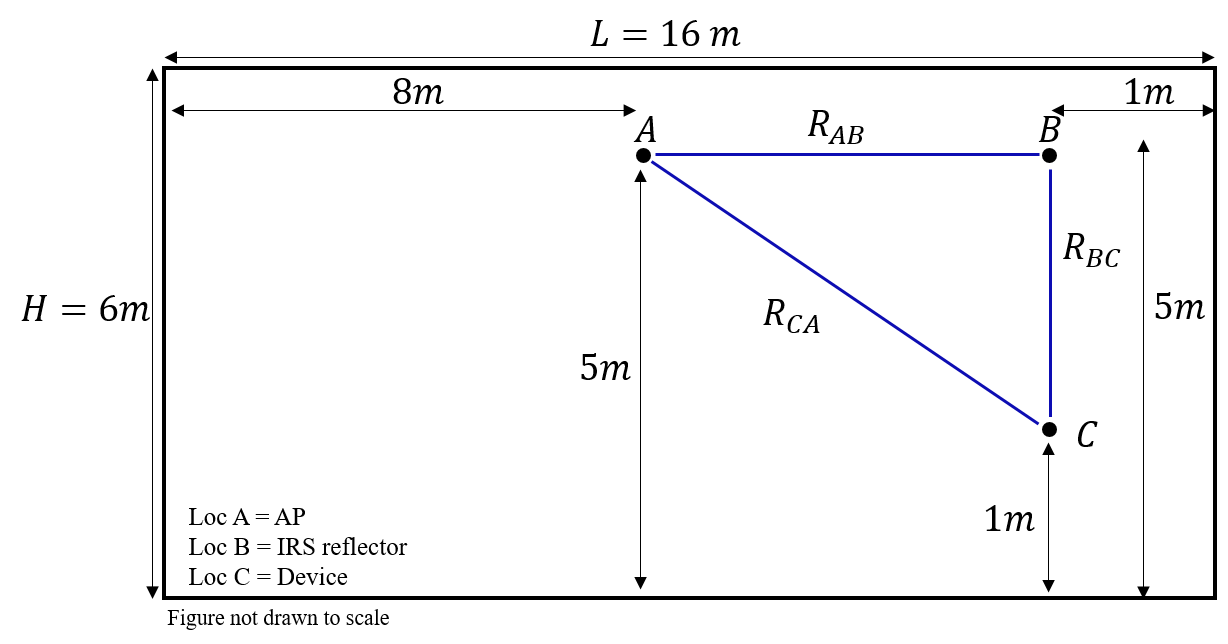}
	\caption{Geometrical representation of elements within PTV used for simulation.}
	\label{fig_2}
    \end{figure}

    The model diagram for the IRS-aided MEC-WET system within PTV is shown in Figure \ref{fig_1}. WE consider a PTV-AP that is capable of transmitting data and energy signals. The MEC server is co-located with the PTV-AP and hence AP-to-server latency is negligible. The IRS' reflecting elements are coated on the vehicle's interior as shown in Figure \ref{fig_1}. A single-antenna PTV-AP and a single device are considered for the simulation to show the effectiveness of the system. Figure \ref{fig_2} shows the geometrical representation of PTV-AP, IRS, and device with default locations and distances. 
    
    A practical non-linear energy harvesting model \cite{zhou2020computation} is considered for the simulation in which the harvesting is successful if the received signal power is above a threshold $P_{th}$, zero otherwise. The harvested power cannot exceed an upper threshold limit of $P_{max}$. Parameters considered in \cite{zhou2020computation} are used in  our simulation as well, unless otherwise stated. 

    The channel model consist of both small and large-scale fading. In particular, exponentially distributed channels with zero mean and unit variance are considered in both downlink and uplink direction. Channels are independent and identically distributed. The IRS placement is considered such that PTV-AP and IRS elements follows LoS link with pathloss parameter $\alpha_o = 2.4$. The pathloss parameter for  PTV-to-User and IRS-to-User channels is $\alpha_n = 3.0$. 
    
    At the start of time-frame of length $\tau$, user device generates $l$ Kbits of data per unit of time, where $l$ is distributed uniformly between 40 Kbits and  50 Kbits. If the device wants to offload its task, it sends a message with relevant information to the PTV-AP which makes decision regarding offloading or local computing. In case of offloading, $l_o = \mu l$ Kbits of data is offloaded to MEC server and $l_c = (1-\mu)l$ Kbits is computed locally by the device, where $\mu \in [0,1]$. Device's CPU takes $\psi = 10^3$ cycles to compute one bit of raw data and energy consumed by CPU to compute $l_c$ bits is $\xi \psi^3 l_c^3$ mJoule, where $\xi = 1\times10^{-28}$ is the effective CPU capacitance \cite{zhou2020computation}. After offloading $l_o$ Kbits, the device saves $\xi \psi^3 l_o^3$ mJoule energy. We considered PTV-AP transmit power $P_t = 1 $ Watt. For the IRS, amplitude reflection coefficient is set to 1 \cite{bai2020latency} and phase shift coefficient ($\Phi$) is set to $\frac{\pi}{2}$.  Each simulation is iterated 1000 times and the averaged results are reported in the following.

    \subsection{Numerical Results} 
     
\begin{figure}[!t]\centering
	\includegraphics[width=\linewidth, trim={0.5cm 0.0cm 1.0cm 0.0cm},clip = true]{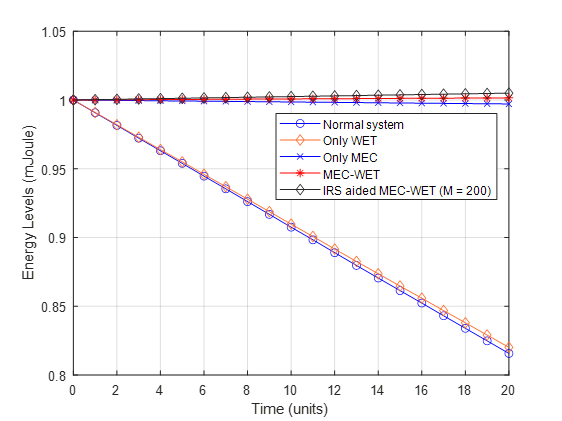}
	\caption{Device’s energy levels with respect to time $P_{max}$= .004 Watts, $\mu$= 0.75 , $\tau_v$= 0.25}
	\label{fig_3}
    \end{figure}

\begin{figure}[!t]\centering
	\includegraphics[width=\linewidth, trim={0cm 0cm 0cm 0.0cm},clip = true]{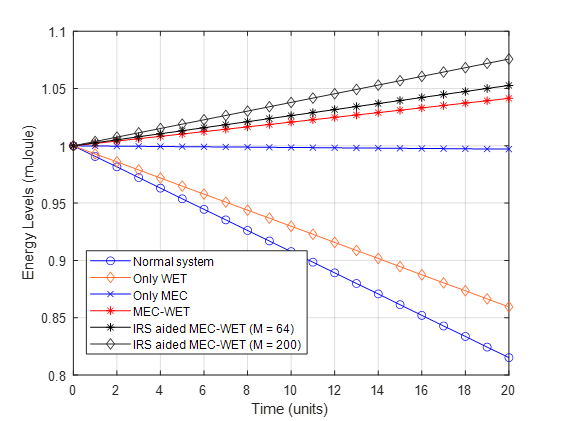}
	\caption{Device’s energy levels with respect to time $P_{max}$= .04 Watts, $\mu$= 0.75 , $\tau_v$= 0.25}
	\label{fig_4}
    \end{figure}

    First we discuss Figure \ref{fig_3} and Figure \ref{fig_4} in which device's energy levels $E[t]$ are shown with respect to time $t$. Device's new energy level in the subsequent time steps is $E[t+1] = E[t] + E_g[t+1] - E_c[t+1] $, where $E_g[t] = E_v[t] + E_{o}[t]$ is the instantaneous energy gain at time $t$ and $E_v[t], E_o[t]$ denote the harvested energy and energy preserved due to offloading, respectively. The energy consumption is calculated as $E_c = E_{ckt} + E_{lc} + E_{tr} $, where $E_{ckt}, E_{lc}, E_{tr}$ denote device circuit's constant energy consumption, energy utilization for local computing, and energy needed to offload task to MEC server. The mobile device's energy level is normalized to 1 mJoule at the start of the simulation (i.e. $t=0$). 
    
    
    Both figures (Fig. \ref{fig_3} and Fig. \ref{fig_4}) show the consumption of device's battery energy levels under normal scenario. There is some energy gain due to WET, however it is not very significant particularly due to poor channel condition between PTV-AP and the device. Even though the IRS aided transmission significantly improves signal strength at the receiving device, the benefits reaped from the reflecting surface are still limited due to device-level saturation \cite{wang2017wirelessly}. This is because in a practical WET system, a device cannot harvest energy beyond its saturation limit due to hardware constraints. To fully exploit the IRS-aided MEC-WET, the maximum allowed harvesting power threshold at the receiving device should be improved. This problem require hardware-level solution from the device manufacturers and is out of scope of theoretical research. As the maximum harvesting power threshold in Figure 4 is $P_{max}= 0.004$ Watts, WET does not offer any significant energy gains. On the other hand, when $P_{max}$ is increased to 0.04 Watts, moderate WET-enabled energy gain is observed in Figure 5, even without IRS.

    Figure \ref{fig_3} and Figure \ref{fig_4} also show that the significant gain in energy occurs due to MEC offloading. For example, for $\mu = 0.75$ and $\tau_0.25$, negligible energy losses are observed for the MEC-only case. This is due to the fact that MEC server takes the toil of computation as result of task offloading which allows energy preservation at user-device end. In case of IRS-aided MEC-WET scenario, the device’s battery gains significantly more energy than that consumed. In fact, device is able to actually store energy above the starting level of 1 mJoule. It can be observed in Figure 5 that by increasing number of reflecting elements from 64 to 200 the gain due to energy harvesting improved considerably. 
  
\begin{figure}[!t]\centering
	\includegraphics[width=\linewidth, trim={0cm 0cm 0cm 0.0cm},clip = true]{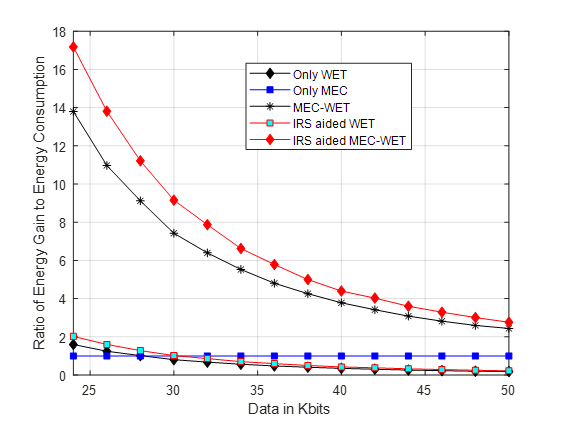}
	\caption{Ratio of energy gain-to-consumption ratio N=64, $P_{max}$ = 0.04 Watt, $N = 64$, $\tau_v = 0.25, \mu = 0.5$.}
	\label{fig_6}
    \end{figure}

    In Figure \ref{fig_6} we show the energy gain to consumption ratio for different amount of data generated by user device. Considering $\mu = 0.5$, we can observe that the ratio for MEC-only scenario and remains below 1 for WET-only and IRS-aided WET scenarios when data size is greater than 30 Kbits. For smaller data size (e.g. 24 Kbits), energy gain in case of IRS-aided WET is twice that of energy consumed by the device. On the other hand, the ratio is significantly higher for MEC and IRS-aided MEC-WET scenario, particularly when data size are low. The ratio decreases as the size of data increases because device need more power for offloading and computing remaining data locally. 
    
    \section{Conclusion}
    
    In this paper we present an idea for the futuristic public transport vehicle that is equipped with in-vehicle AP, Mobile Edge Computing and Wireless Energy Transfer facilities and use Intelligent Reflecting Surface to improve in-vehicle channel condition. Since preserving device battery levels for commuting users is a serious challenge, future F-PTVs will replenish commuters’ smart devices by employing IRS aided MEC-WET. The simulation results for IRS-aided integrated MEC-WET system showed significantly increased device battery’s energy levels. Some challenges associated with the proposed F-PTV and their solutions are also presented in this paper. Such a model is still open for research. As our future work, we are working on jointly optimizing $\tau_v,\mu$ and phase shift coefficient $\Phi$ within the proposed framework, along with designing MEC-WET resource scheduling scheme for  multiple user scenario.
     
    \section{Acknowledgment}
    This work is submitted to IEEE Vehicular Technology Magazine for publication. It is supported by XJTLU's Research Development Fund: RDF-20-01-15.

\bibliographystyle{IEEEtran}
\bibliography{sixthGen}


  

\end{document}